\documentclass[aps,prl,floatfix,twocolumn,showpacs,superscriptaddress]{revtex4}

\usepackage{graphicx,epsfig}% Include figure files
\usepackage{times}
\usepackage{graphics,dcolumn,bm,fleqn,epic,eepic,float}
\usepackage{amssymb,amsmath,multirow,rotate,color}
\begin{document}

\title{Annealed and Mean-Field formulations of Disease Dynamics on Static
  and Adaptive Networks}

\author{Beniamino Guerra}

\affiliation{Laboratorio sui Sistemi Complessi, Scuola Superiore di
  Catania, Universit\`a di Catania, 95123 Catania, Italy}

\affiliation{Departamento de Matem\'atica Aplicada, Universidad Rey Juan 
Carlos (ESCET), 28933 M\'ostoles (Madrid), Spain}

\author{Jes\'us G\'omez-Garde\~nes}

\affiliation{Departamento de Matem\'atica Aplicada, Universidad Rey Juan 
Carlos (ESCET), 28933 M\'ostoles (Madrid), Spain}

\affiliation{Institute for Biocomputation and Physics of Complex
Systems (BIFI), University of Zaragoza, 50009 Zaragoza, Spain}

\date{\today}

\begin{abstract}
We use the annealed formulation of complex networks to study the
dynamical behavior of disease spreading on both static and adaptive
networked systems. This unifying approach relies on the annealed
adjacency matrix, representing one network ensemble, and allows to
solve the dynamical evolution of the whole network ensemble all at
once. Our results accurately reproduce those obtained by extensive
numerical simulations showing a large improvement with respect to the
usual heterogeneous mean-field formulation. Moreover, by means of the
annealed formulation we derive a new heterogeneous mean-field
formulation that correctly reproduces the epidemic dynamics.
\end{abstract}

\pacs{89.75.Fb, 02.50.Ga, 05.45.-a} 
\maketitle

%____________________________________________________________________
%                           INTRODUCTION
%____________________________________________________________________

A variety of natural and socioeconomic complex systems have a
networked interaction backbone. This interaction backbone is usually
described as a graph where nodes represent the constituents of the
system and edges account for the relation between them. The complex
network's approach has proved to be a powerful tool to unveil common
topological properties of systems related to seemingly different
fields \cite{rev:albert}. The ubiquity of properties such as the
small-world phenomenon and the scale-free character of real
interaction networks, has spurred on applied and theoretical research
aimed at understanding the origin of their underlying principles.

%Some examples of the intricate relation between
%structure and function include the spread of information and diseases
%in societies \cite{rev:doro2}, the survival of cooperation and the
%culture diffusion in socioeconomical systems
%\cite{rev:anxo,rev:castellano}, or the emergence of synchrony in
%coupled biological and chemical systems \cite{rev:arenas}.

In the last decade the studied on how structural properties of complex
networks affect their functionality \cite{rev:bocc,rev:doro2} have
focused the attention of the physics of complex systems. Most of the
theoretical studies about the dynamics on top of complex networks make
use of synthetic networks with some prescribed structural properties
in order to study the impact that these topological features have on
the dynamical behavior. Typically, the dynamics is carried on top of a
number of different networks that can be seen as microstates of a
large network ensemble characterized by some topological properties of
interest. The numerical simulation of the dynamics on top of a large
enough number of network realizations belonging to the same
topological ensemble allows to obtain meaningful averages for the
dynamical quantities that characterize the state of the
system. However, the computational costs increase with the complexity
of both the network ensemble and the dynamics at work.
Apart from numerical simulations, dynamical processes on static
networks have been widely studied by means of the heterogeneous
mean-field (HMF) approximation \cite{HMF1,HMF2,HMF3}. This formulation
relies on the dynamical equivalence of the nodes belonging to the same
degree class, {\em i.e.} the set of nodes with the same number of
neighbors. The HMF coarse-graining has provided important insights
about the critical phenomena taking place on several dynamical
processes \cite{rev:doro2} and, although it has been recently debated
in the context of contact processes \cite{PRL06,PRL07c,PRL07r,PRL08},
it has constituted the main theoretical framework for the study of
dynamical processes on complex networks.

In this Letter we study the Susceptible-Infected-Susceptible (SIS)
epidemic dynamics by means of the annealed adjacency matrix (AAM) of
an entire network ensemble. We show how, inserting the AAM into the
microscopic equations for the dynamics of the nodes, it is possible to
overcome the need of statistics over network realizations and to
obtain accurate results about the dynamical state of the system. On
the other hand, we show that the HMF formulation fails to reproduce
the SIS diagram, even for the case of uncorrelated networks. Moreover,
we will use the annealed formulation to derive a new HMF formulation
that, for the first time, captures the entire epidemic phase
diagram. Finally, we show how the annealed formalism can be
efficiently applied to the important case of adaptive networks in
which network growth coevolves with SIS dynamics, thus providing with
an unifying framework for the study of general epidemic dynamics in
complex systems.

% In particular, we will make use of the annealed formulation to
%describe growing systems where the addition of new nodes is coupled
%with the dynamical state of the units. We will apply the method to
%the spread of susceptible-infected-susceptible (SIS) disease on both
%static and growing networks and show that our results nicely agree
%with those obtained carrying costly numerical simulations.

%____________________________________________________________________
%                           STATIC NETWORKS
%____________________________________________________________________

The idea behind the annealed approximation is to deal directly with
the network ensemble of interest rather than with a collection of
network realizations (see
\cite{Bianconi02,Bianconi08a,Bianconi08b,PRE09,PRL09-1,PRL09-2} for
its application to the study of both structural and dynamical
properties of networks). A ensemble is a family of networks, each of
them represented by an adjacency matrix, $A$, with binary entries:
$A_{ij}=1$ when nodes $i$ and $j$ are connected and $A_{ij}=0$
otherwise. All the networks belonging to the same ensemble share some
topological quantities, being the most general the ensemble of graphs
with fixed number of nodes, $N$, and links, $L$. Other ensembles
recently explored \cite{Bianconi08a} are those composed of networks
with the same degree distribution $P(k)$, {\em i.e.} the probability
of having a node with $k$ connections. One network ensemble can be
described by a single matrix, the AAM, whose terms, ${\mathcal
  A}_{ij}$, account for the probability that two nodes share a
connection as dictated by the ensemble constraints. To construct the
AAM of a given network ensemble, one can average over the set of
adjacency matrices or to define each element ${\mathcal A}_{ij}$ from
scratch by calculating the probability of attaching $i$ to $j$. In the
following we will focus on ensembles of undirected networks,
${\mathcal A}_{ij}={\mathcal A}_{ji}$.

To show the use of the annealed approximation for the study of
dynamical processes on top of static networks we will analyze the
spreading of a SIS disease. The SIS disease spreading takes place as
usual: Starting from a fraction $I_0$ of infected nodes, at each time
step the infected nodes attempt to infect each of their neighbors with
a success probability $\lambda$. Additionally, infected nodes recover
(and become susceptible of being infected in the next time steps) with
probability $\mu$. Typically, after enough time steps the number of
infected individuals reach a steady number that characterizes the
impact of the disease. The infection probability $\lambda$ plays a key
role on the disease spreading. In particular, when the ratio
$\lambda/\mu$ exceeds some critical value $(\lambda/\mu)_c$ the
disease infects a macroscopic part of the population whereas below
this epidemic threshold the time evolution of the number of infected
nodes vanishes. Therefore, $\lambda/\mu$ acts as a control parameter
of the phase transition from the healthy to the epidemic phase while
the asymptotic fraction of infected nodes, $I$, is the order
parameter.

The SIS phase diagram of a network ensemble described by a given AAM
can be accurately obtained by using a Markov chain formulation similar
to that introduced in \cite{Markov} for particular network
realizations. In our case, we denote as $s_{i}(t)$ the probability
of finding node $i$ in the healthy state ($s_{i}=1$) after $t$ time
steps in a randomly chosen network of the ensemble. We can express the
evolution for the set of probabilities $\{s_{i}(t)\}$ through the
following set of discrete-time Markovian equations:
\begin{equation}
s_{i}(t+1)= s_{i}(t)+\mu\left[1-s_{i}(t)\right]-s_{i}(t)
\left[1-q_{i}(t)\right]\;,
\label{eq:Markov}
\end{equation}
where $q_{i}(t)$ is the probability that node $i$ is not infected by
any neighbor:
\begin{equation}
q_{i}(t)=\prod_{j=1}^{N}\left[1-\lambda {\mathcal A}_{ij}(1-s_{j}(t))\right]\;.
\label{eq:Markov2}
\end{equation}
The last two terms of the right-hand side of equation
(\ref{eq:Markov}) correspond to the recovery of infected nodes and the
infection of healthy nodes respectively.  Note, that the above time
evolution has the same functional form of the equations introduced in
\cite{Markov} but, instead of having a particular adjacency matrix
$A_{ij}$, we have the probability that two nodes, $i$ and $j$, are
connected in a randomly chosen network of the ensemble, {\em i.e.}
${\mathcal A}_{ij}$. By iterating equations (\ref{eq:Markov}) from the
initial condition $s_{i}=1-I_{0}$ $\forall i$, a stationary
distribution $\{s^{\infty}_{i}\}$ is reached and the order parameter
$I$ for the corresponding network ensemble is computed as
$I=\sum_{i}(1-s^{\infty}_{i})/N$.

%CONFIGURATIONAL

\begin{figure}[t!]
\epsfig{file=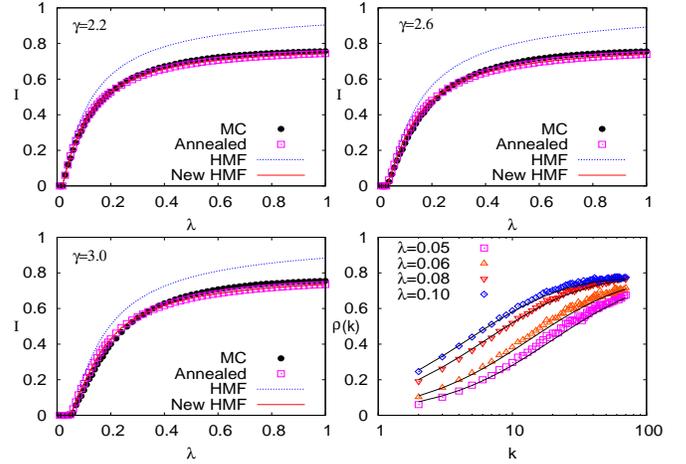,width=3.5in,angle=0,clip=1}
\caption{The top and bottom left panels show the phase diagrams,
  $I(\lambda)$, for the numerical simulations (points), the annealed
  formalism (squares), the usual (dashed line) and the new (solid
  line) HMF approximation. The networks have been constructed using
  the configurational model as introduced in \cite{PRE05} with degree
  sequences following power-law degree distributions, $P(k)\sim
  k^{-\gamma}$, with $\gamma=2.2$, $2.6$ and $3.0$. The networks have
  $N=5000$ and $\langle k\rangle=4$ and the statistics of the
  numerical simulations is made over $500$ network realizations for
  each value of $\lambda$.  Besides, the bottom right panel show the
  fraction of infected nodes of degree $k$, $\rho(k)$, using numerical
  simulations (points) and the annealed formalism for the case
  $\gamma=2.2$ and several values of $\lambda$. The recovery
  probability is set to $\mu=0.3$.}
\label{fig:1}
\end{figure}

The most general network ensemble is provided by the configurational
model \cite{configurational} that generate a family of graphs by
specifying a fixed degree sequence for the $N$ nodes, $\{k_{1},
k_{2},...,k_{N}\}$. Each particular network realization is constructed
by sorting each of the $L$ available links ($L=\sum_{i}k_{i}/2$). The
probability that two nodes $i$ and $j$ are chosen to be connected in
one network realization depends on their respective degrees as
$k_{i}k_{j}/(2L^2)$. Therefore, the probability that, after sorting
the $L$ links, two given nodes $i$ and $j$ with degrees $k_{i}$ and
$k_{j}$ are connected is ${\mathcal A}_{ij}={\mathcal
  A}_{ij}=k_{i}k_{j}/(2L)$. Now, inserting this latter AAM in
eqs. \ref{eq:Markov} and \ref{eq:Markov2} we can study the SIS
dynamics of configurational ensembles.
In Fig. \ref{fig:1} we represent the curves $I(\lambda)$ and the
fraction of infected individuals with degree $k$, $\rho (k)$ obtained
via the annealed formulation and the use of extensive numerical
simulations of the SIS dynamics. The results of $I(\lambda)$ are shown
for three families of configurational ensembles corresponding to
$P(k)\sim k^{\gamma}$ with $\gamma=2.2$, $2.6$ and $3.0$. The network
realizations used for the numerical simulations were constructed using
the method introduced in \cite{PRE05} so to assure that no
degree-degree correlations are present in any of the networks
generated. As shown in Fig. \ref{fig:1}, the results obtained using
the AAM strongly agree with those obtained through extensive numerical
simulations.

We have compared the results of the AAM formulation with those
obtained with the HMF. Surprisingly, the HMF fails to reproduce the
phase diagrams $I(\lambda)$ as observed in Fig. \ref{fig:1}. The
reason of these discrepancies can be explained with the annealed
formulation of the SIS dynamics. Taking the continuous-time
formulation and assuming that all the nodes with the same degree have
the same probability of being healthy, {\em i.e.}  $s_{i}=\rho_{k}$
$\forall i$ with $k_{i}=k$, we can derive a new HMF approximation from
eqs. (\ref{eq:Markov}) and the expression of ${\mathcal A}_{ij}$:
\begin{equation}
\dot{\rho}_{k}=\mu\left(1-\rho_{k}\right)-\rho_{k}\left[1-\prod_{k^{'}}\left[1-\lambda
    \frac{kk^{'}}{2L}(1-\rho_{k^{'}})\right]^{N_{k^{'}}}\right]
\label{eq:newHMF1}
\end{equation}
where $N_{k}=NP(k)$. This new HMF approximation when integrated for
the different model networks perfectly reproduces the annealed
approximation in contrast with the usual HMF approximation extensively
used in network epidemiology up to date \cite{HMF1,HMF2,HMF3}.

The natural extension of eqs. (\ref{eq:newHMF1}) and
(\ref{eq:newHMF2}) to general network models is done by considering
that the probability of finding two nodes of degrees $k$ and $k^{'}$
in the configurational model, $P(k^{'},k)=kk^{'}/2L$, reads as
$P(k^{'},k)=kP(k^{'}|k)/(NP(k'))$ for general network models. Then, we
can write the new HMF as:
\begin{equation}
\dot{\rho}_{k}=\mu\left(1-\rho_{k}\right)-\rho_{k}\left[1-\prod_{k^{'}}\left[1-\lambda
    \frac{kP(k^{'}|k)}{N_{k^{'}}}(1-\rho_{k^{'}})\right]^{N_{k^{'}}}\right]
\label{eq:newHMF2}
\end{equation}
Note that the new HMF formulation takes the form of the usual HMF
equations only when $\lambda(1-\rho_{k})\ll 1$ $\forall k$:
\begin{equation}
\dot{\rho}_{k}\simeq\mu\left(1-\rho_{k}\right)-\lambda k\rho_{k}\sum_{k^{'}}P(k^{'}|k)(1-\rho_{k^{'}})+{\cal O}(\lambda^2)
\label{eq:oldHMF}
\end{equation}
Therefore, the usual HMF is only correct for $\lambda$ values before
  and around the epidemic onset (as observed from Fig. \ref{fig:1}).

%ATTACHMENT

Now we describe the case of ensembles of networks constructed via
attachment processes. In this case, network realizations are usually
assembled by starting (at $\tau=0$) from a small graph of $m_{0}$
nodes coupled all-to-all. Therefore, the AAM has ${\mathcal A}_{ij}=1$
for $i$ and $j=1$, ..., $m_{0}$ provided $i\neq j$ and ${\mathcal
  A}_{ii}=0$ otherwise. At each step $\tau$ of the network growth
($\tau=1$,...,$N-m_{0}$) a new node is added and launches $l$ links to
the nodes of the network. Each of these nodes have some attachment
probability, $\Pi_{i}(\tau)$ ($i=1$,...,$\tau+m_{0}-1$), of receiving
one of the $l$ links from the new node. Therefore, the remaining
$(N^2-m_{0}^2)$ terms of the AAM read ${\mathcal
  A}_{ij}=1-(1-\Pi_{m}(M))^{l}$ (where $m=\min(i,j)$ and
$M=\max(i,j)$) when $i\neq j$, and ${\mathcal A}_{ii}=0$
otherwise. Once the AAM is constructed one can obtain the annealed
degree of the nodes as $k_{i}=\sum_{j}{\mathcal A}_{ij}$.

\begin{figure}[t!]
\epsfig{file=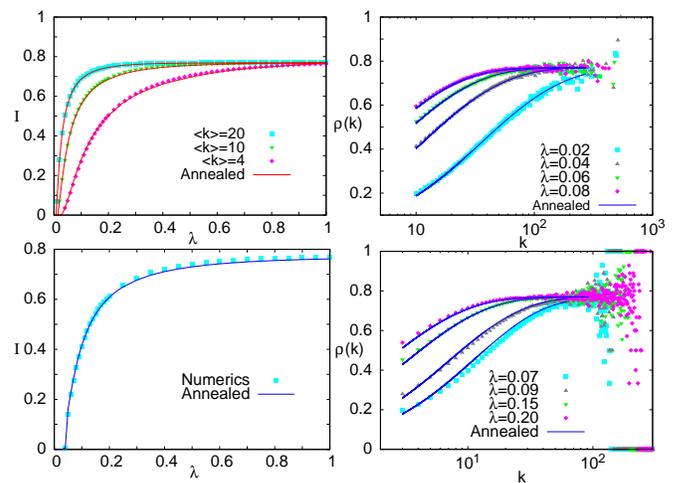,width=3.5in,angle=0,clip=1}
\caption{(Top) In the left we show the epidemic phase diagram
  $I(\lambda)$ for three BA ensembles with $\langle k\rangle=4$, $10$
  and $20$. In the right we show the fraction of infected individuals
  across degree-classes $\rho(k)$ for the ensemble with $\langle
  k\rangle=4$ and for different values of $\lambda$. The points
  correspond to numerical simulations of the SIS dynamics on top of
  $500$ network realizations while curves stand for the annealed
  approximation. The size of the networks is $N=5000$. (Bottom) Again,
  we show both $I(\lambda)$ (left) and $\rho(k)$ (right) for the case
  of adaptive growing networks when $N=10^4$. Both results of
  numerically grown networks (points), averaged over $5000$
  realizations, and the solutions of the annealed formalism (curves)
  are shown. The network development is characterized by $\mu=0.3$,
  $\tau_{T}=1$ and $l=3$.}
\label{fig:2}
\end{figure}

The networks grown through attachment processes are know to have
nontrivial node-to-node correlations ({\em e.g.} age correlations)
that play a key role in their dynamical behavior and are difficult to
capture via HMF formulations. Now we show that these peculiarities of
the attachment ensembles are captured by the annealed formulation of
the SIS dynamics. Let us focus on the preferential attachment kernel
introduced by Barab\'asi and Albert (BA),
$\Pi_{i}(\tau)=k_{i}(\tau)/\sum_{j}k_{j}(\tau)$ \cite{BA}. In this
case the networks have a power law degree distribution, $P(k)\sim
k^{-3}$, and average degree $\langle k\rangle=2m$. Thus, by fixing the
value of $m$ we can construct an ensemble of BA networks with the same
values of $\langle k\rangle$. In Fig. \ref{fig:2} we show the results
of $I(\lambda)$ and $\rho(k)$ by solving equations \ref{eq:Markov}
with the expression of ${\cal A}_{ij}$ corresponding to three BA
ensembles with $m=2$, $5$ and $10$. Again, the results of the AAM
approximation show a perfect agreement with those obtained by means of
numerical simulations of the SIS dynamics on top of BA networks.

%____________________________________________________________________
%                           ADAPTIVE NETWORKS
%____________________________________________________________________

%adaptive growing networks. In this general case, networks
%grow simultaneously to the internal dynamics of the system
%\cite{PLoS1,NJP}. These latter processes shape the structure of the
%system thus closing the feedback loop between system's form and
%function. The two entangled processes (growth and system's function)
%coevolve in such a way that the attachment preferences of newcomers
%are driven by the dynamical states of their predecessors. To this
%purpose, 

Now we focus on a more complicated attachment kernel describing a
network growth interplaying with the internal dynamics of the
nodes. In particular, the two entangled processes (network growth and
system's dynamics) coevolve in such a way that the attachment
preferences of newcomers are driven by the dynamical states of their
predecessors. Therefore, the attachment probability of a node $i$ at
time $t>i$, $\Pi_{i}(t)$ depends, not only on the topological
properties of node $i$, but also on its instant dynamical state
$x_{i}(t)$. In our case $x_{i}(t)=s_{i}(t)$ and the attachment kernel
reads: $\Pi_{i}(t)=k_{i}(t)s_{i}(t)/\sum_{j}k_{j}(t)s_{j}(t)$.
%We introduce an adaptive growing system where network grows together
%with the spread of a SIS disease. Starting from an all-to-all
%connected core of $m_{0}$ nodes we inoculate the disease into a
%fraction $I_{0}$ of them. At each time step, the SIS disease spreads
%across the population as described for static networks. Additionally,
%at each $\tau_{T}$ time steps a new healthy node is incorporated into
%the system and connects to $l$ nodes. 
%Network growth and SIS dynamics
%are coupled via the attachment kernel so that Thus, the attachment kernel yields:
This particular kernel couples networks growth and SIS dynamics so
that newcomers prefer acquaintances with both structural importance
and a healthy state. Therefore, infected nodes cannot receive a link
from newcomers until they recover, whereas those healthy elements
compete with each other according to their degrees as in the
preferential attachment model.

The study of the above adaptive growth mechanism can be efficiently
tackled with the help of the annealed formulation. To this aim we
consider again the Markov SIS equations (\ref{eq:Markov})
incorporating a time-dependent AAM, ${\mathcal A}(t)$, in equation
(\ref{eq:Markov2}) that evolves simultaneously to the disease
spreading. The initial condition of the AAM is set to ${\mathcal
  A}_{ij}(0)=1$ for $i,j\leq m_{0}$, $i\ne j$ and ${\mathcal
  A}_{ij}=0$ otherwise, while the dynamical state of the nodes is
initialized as: $s_{i}(0)=1-I_{0}$ when $i\leq m_{0}$ and $s_{i}(0)=1$
otherwise. Then, we couple equations (\ref{eq:Markov}) with the
evolution of the AAM:
\begin{equation}
{\mathcal A}_{ij}(t)=H(t-\tau_{T}M)
\left[1-(1-\Pi_{m}(\tau_{T}M))^{l}\right]\;,
\label{AAMattach}
\end{equation}
when $i\neq j$ and $A_{ii}=0$ otherwise. The function $H(x)$ is the
Heaviside step function so that the above equations state that the
element ${\mathcal A}_{ij}$ has a zero value until the youngest
element $M$ is incorporated. At this step, $t=\tau_TM$, the value of
${\mathcal A}_{ij}$ jumps to a stationary value that only depends on
the attachment probability of the oldest node $m$ at time $M$.

\begin{figure}[t!]
\epsfig{file=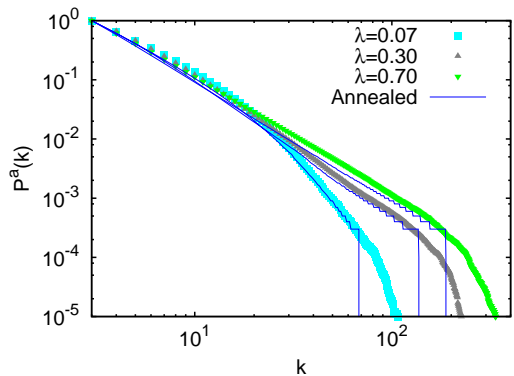,width=2.0in,angle=-90,clip=1}
\caption{Accumulated degree distribution of growing adaptive networks
  when $N=10^4$. The numerical results (points) and the distribution
  obtained from the evolution of the AAM (curves) are shown for
  several values of $\lambda$. The networks are grown using
  $\tau_{T}=1$, $l=3$ and $\mu=0.3$. Numerical results are obtained
  averaging over $100$ network realizations for each value of
  $\lambda$.}
\label{fig:3}
\end{figure}

Iterating the above dynamics until the system size have reached a
large enough value $N$ (so that it has reached a stationary regime for
its intensive observables) we can evaluate the asymptotic impact of
the disease and the structural properties of the generated
networks. First, in Fig. \ref{fig:2} we show the phase diagram
$I(\lambda)$ and the microscopic distribution of infected nodes across
degree-classes, $\rho(k)$. Both measures clearly show the accuracy of
the annealed approximation when compared to the results obtained
averaging over a number of numerically grown networks. Besides, in
Fig. \ref{fig:3} we show the corresponding accumulated degree
distributions of the grown networks using different infection
probabilities $\lambda$. From the figure we show that the AAM
generated reproduces correctly the degree distributions obtained
numerically. In particular, it is shown that for small values of
$\lambda$ rather homogeneous networks are obtained, pointing out that
the prevalence of disease on network hubs, as shown in
Fig. \ref{fig:2} (bottom right panel), screens their fitness to
attract new links and thus avoids the development of degree
heterogeneity. On the contrary, for large $\lambda$ the disease
affects homogeneously the forming system and newcomers mainly guide
their links towards large degree nodes, thus recovering the behavior
of the BA preferential attachment model.

%__________________________________________________________________
%                           CONCLUSIONS
%___________________________________________________________________

Summing up, in this Letter we have shown that the dynamical behavior
of static and adaptive networks can be efficiently described by the
use of an unifying framework: the annealed formulation of the
adjacency matrix. In both cases, the annealed formulation allows to
compute the macroscopic observables and the microscopic distributions
of the dynamics of a network ensemble in one single computation, thus
avoiding the need of extensive simulations.
In the case of adaptive networks, the treatment of the AAM as a
dynamical object allows a simple formulation of the coupled
structure-function problem. Particularizing to the case of SIS disease
spreading on top of static networks, we have shown that the HMF
approximation does not reproduces the phase diagram. Moreover, using
the annealed formulation of the dynamics we have obtained a new HMF
formulation that shows, for the first time, an accurate agreement with
the numerical results.
The use of the annealed formulation in other kind of network dynamics
only requires to substitute the usual adjacency matrix by the AAM in
the microscopic equations at work. Therefore, the formalism allows to
write down explicitly the evolution equations for a wide variety of
network ensembles, adaptive schemes and dynamics, regardless of their
stochastic ({\em e.g.}  contact processes, information diffusion,
evolutionary dynamics, etc) or deterministic ({\em e.g.}  nonlinear
dynamical systems) nature.
Although more research is needed to assess to what extent this
approach can be such an important improvement in other dynamic network
models, we expect that the annealed formulation will allow to tackle
several open problems on networks dynamics and will pave the way to
the theoretical studies of the growing field of complex adaptive
systems.

%____________________________________________________________________

\begin{acknowledgments}
 We thank V. Latora, Y. Moreno and A. S\'anchez for useful
 comments. This work has been supported by MICINN through Grants
 FIS2008-01240 and MTM2009-13848.

\end{acknowledgments}


\begin{thebibliography}{99}

\bibitem{rev:albert} R. Albert and A.-L. Barab\'asi,
  Rev. Mod. Phys. {\bf 74}, 47 (2002).

\bibitem{rev:bocc} S. Boccaletti, V. Latora, Y. Moreno,
  M. Chavez and D.-U. Hwang, Phys. Rep. {\bf 424}, 175 (2006).

\bibitem{rev:doro2} S.N. Dorogovtsev, A.V. Goltsev, and J.F.F. Mendes,
  Rev Mod. Phys. {\bf 80}, 1275 (2008).

\bibitem{HMF1} R. Pastor-Satorras and A. Vespignani,
  Phys. Rev. Lett. {\bf 86}, 3200 (2001).

\bibitem{HMF2} A.L. LLoyd and R.M. May, Science {\bf 292}, 1316 (2001).

\bibitem{HMF3} Y. Moreno, R. Pastor-Satorras and A. Vespignani
  Eur. Phys. J. B {\bf 26}, 521 (2002).

\bibitem{PRL06} C. Castellano and R. Pastor-Satorras,
  Phys. Rev. Lett. {\bf 96}, 038701 (2006).

\bibitem{PRL07c} M. Ha, H. Hong and H. Park, Phys. Rev. Lett. {\bf 98}, 029801 (2007).

\bibitem{PRL07r} C. Castellano and R. Pastor-Satorras,
  Phys. Rev. Lett. {\bf 98}, 029802 (2007).

\bibitem{PRL08} C. Castellano and R. Pastor-Satorras,
  Phys. Rev. Lett. {\bf 100}, 148701 (2008).

\bibitem{Bianconi02} G. Bianconi, Phys. Lett. A. {\bf 303}, 166 (2002).

\bibitem{Bianconi08a} G. Bianconi, A.C.C. Coolen and C. J. Perez
  Vicente, Phys. Rev. E {\bf 78}, 016114 (2008).

\bibitem{Bianconi08b} G. Bianconi, EPL {\bf 81}, 28005 (2008).

\bibitem{PRL09-1} S. Carmi, S. Carter, J. Sun and D. ben-Avraham,
  Phys.  Rev. Lett. {\bf 102}, 238702 (2009).

\bibitem{PRL09-2} C. Caretta Cartozo and P. De Los Rios,
  Phys. Rev. Lett.  {\bf 102}, 238703 (2009).

\bibitem{PRE09}  M. Bogu\~n\'a, C. Castellano and R. Pastor-Satorras, Phys. Rev. E {\bf 79}, 036110 (2009).

\bibitem{Markov} S. G\'omez, A. Arenas, J. Borge-Holthoefer, S. Meloni
  and Y. Moreno, EPL {\bf 89}, 38009 (2010).

\bibitem{configurational} M. Molloy and B. Reed. Comb. Prob. and Comp. {\bf 7}, 295 (1998). 

\bibitem{PRE05} M. Catanzaro, M. Bogu\~na and R. Pastor-Satorras,
  Phys. Rev. E {\bf 71}, 027103 (2005).

%\bibitem{PNAS08} J. G\'omez-Garde\~nes, Y. Moreno, V. Latora and
% E. Profumo, Proc. Nat. Acad. Sci. (USA) {\bf 105}, 1399 (2008).

%\bibitem{PNAS09} S. Meloni, A. Arenas and Y. Moreno,
% Proc. Nat. Acad. Sci. (USA) {\bf 106}, 16897 (2009).

\bibitem{BA} A.L. Barab\'asi and R. Albert, Science {\bf 286}, 509
  (1999).

\end{thebibliography}
\end{document}